# Enhancing Critical Current Density in Thin Superconductor Layers by Moiré Pinning Centers

Z. Owjifard[1], A. Tavana[1*], and M. Hosseini[2*]

[1]AMDM Lab., Department of Physics, University of Mohaghegh Ardabili, Ardabil, Iran.

[2] Department of Physics, Shiraz University of Technology, Shiraz, Iran.

[*]Corresponding author

## Abstract

One important factor affecting the critical current density in type-II superconductors is the formation of artificial pinning centers. Hence, the engineering of pinning centers in superconducting systems has garnered considerable attention. In this study, the effect of moiré patterned pinning centers on the critical current density of superconducting tapes is investigated. The Langevin equation is solved by taking into account the prominent forces within the superconductor medium, using the appropriate boundary conditions for vortices. The vortices dynamics are investigated by performing molecular dynamics simulations, which are used to calculate the corresponding critical current densities. Results show a significant enhancement in the critical current density at particular angles of the relative rotation of the primary lattices. It is also revealed that for stronger pinning forces, the calculated critical current densities are higher in the moiré lattices compared to the primary lattices of pinning centers.

## I- Introduction

Superconductors have attracted considerable attention due to their remarkable ability to conduct electric current without any energy dissipation [1,2]. However, they are more than just perfect conductors and repel magnetic flux, regardless of their history, i.e., the Meissner effect. In type-II superconductors, when specimens are subjected to an external magnetic field lower than $B_{c2}$, they exhibit a characteristic phase known as the vortex or Abrikosov phase, where magnetic vortices partially penetrate the material [3,4]. These vortices organize themselves into a hexagonal triangular lattice structure, commonly referred to as the Shubnikov phase [4,5].

Indeed, the motion of vortices leads to the dissipation of energy, due to the displacement of the normal-phase vortex cores. Therefore, restricting the mobility of vortices is essential in improving the performance of superconductors, as it results in an increase in the dissipation-free current carrying capacity [6]. Hence, studying the dynamics of vortices in the presence of the pinning centers on various types of patterned substrates is crucial in type-II superconductors [7].

Numerous investigations have successfully studied the effect of various geometries of pinning center patterns in superconductor tapes, including triangular [8-23], square [8,9,16,24-37], rectangular [38-43], conformal crystal [44, 45], Kagome [11, 15-17, 21, 46-52], honeycomb [11,19, 46-48,51], and other [53-55] arrays of pinning centers. In practice, these different geometries have been achieved through meticulous experimental techniques and have provided valuable insights into the behavior of vortices in superconducting materials. Through the manipulation of these pinning center patterns, researchers aim to advance the comprehension of vortex dynamics, improve the existing applications, and explore new applications for superconductors.

One class of pinning center arrangements can be obtained from the moiré patterns. The moiré patterns emerge from the interference effects between two overlaid lattices when one of them is shifted or twisted with respect to the other one [56,57]. These patterns can occur on a large scale and are often observed when two regular patterns, such as grids or lines, are combined. They have found interesting applications in various areas, including optics, imaging, and twistronics [56,58].

In this study, the effect of the twist angle of moiré patterns, resulting from the rotation of two triangular pinning center lattices, on the dynamics of vortex lines is investigated. We calculate the critical current density for a range of twist angles $\theta$, and for different strengths of the pinning force, based on molecular dynamics simulations. Results indicate that for certain optimum values of $\theta$, there is a significant enhancement in the critical current density.

### I- Simulation details

Calculations are performed based on the molecular dynamics method. The system comprises magnetic vortices interacting with each other and with the pinning center situated on a moiré pattern. Then, the overdamped equation of motion governing the movement of the *i*-th vortex is [9,30,46, 59-62]:

$$\eta v_i = f_i = f_i^{vv} + f_i^{vp} + f_i^{T} \tag{1}$$

where $\eta$, is the damping coefficient or viscosity and $f_i$ represents the total force acting on the $i$-th vortex. In this equation, $f_i^{vv}$ is the vortex-vortex interaction force that can be expressed as [11,30,46, 59-63]:

$$f_i^{vv} = f_0 K_1\left(\frac{|\mathbf{r}_i - \mathbf{r}_j|}{\lambda}\right) \hat{\mathbf{r}}_{ij}, \tag{2}$$

where, $f_0$ represents the amplitude of the vortex-vortex force, $K_1$ is the modified Bessel function, $\lambda$ is the magnetic field penetration depth, and $r_i$ is the position of the $i$-th vortex, such that [30,46, 59-62]:

$$\hat{\mathbf{r}}_{ij} = \frac{(\mathbf{r}_i - \mathbf{r}_j)}{|\mathbf{r}_i - \mathbf{r}_j|}, \tag{3}$$

$$\phi_0 = \frac{hc}{2e} \tag{4}$$

and

$$f_0 = \frac{\phi_0^2}{8\pi^2 \lambda^3}. \tag{5}$$

In equation (4), $\varphi_0 = hc/2e$ is the quantum of magnetic flux, where h is Planck's constant, $e$ is the electron charge, and $c$ is the velocity of light.

In equation (1), $f_i^{vp}$ is the pinning force described as [64, 66]:

$$f_i^{vp} = \left(\frac{f_p}{r_p}\right) |\mathbf{r}_i - \mathbf{r}_k^{(p)}| \Theta\left(\frac{r_p - |\mathbf{r}_i - \mathbf{r}_k^{(p)}|}{\lambda}\right) \hat{\mathbf{r}}_{ik}^{(p)}, \tag{6}$$

where $f_p$ is the amplitude of the pinning force, $r_p$ is the effective range of the pinning potential, and $\Theta$ is the Heaviside step function. $\hat{\mathbf{r}}_{ik}^{(p)}$ is defined as:

$$\hat{\mathbf{r}}_{ik}^{(p)} = (\mathbf{r}_i - \mathbf{r}_k^{(p)})/|\mathbf{r}_i - \mathbf{r}_k^{(p)}| \tag{7}$$

The lateral dimensions of the system are $L \times L$, where $L$ is set equal to $10\lambda$. Periodic boundary conditions are assumed for both $x$ and $y$ directions. Without loss of generality, we also set $\eta=1$ and $\lambda=1$.

$f_i^T$ in equation (1), is the force resulting from the thermal fluctuations and is treated as a random function that meets the following conditions [67]:

$$<f_i^T(t)> = 0 \tag{8}$$

$$<f_i^t(t) f_i^t(t')> = 2\eta k_B T \delta_{ij} \delta(t-t') \tag{9}$$

In this approximation, the normalized critical current density, $J_c$, is defined as the ratio of the number of pinned vortices, $N^p{}_v$ to the total number of vortices, $N_v$, [64,65] i.e.:

$$J_c(\phi) \propto \frac{N_v^p(\phi)}{N_v(\phi)}, \tag{10}$$

For the geometry of the pinning centers, the initial configuration comprises two triangular lattices placed exactly on top of each other with a relative twist angle equal to zero ($\theta = 0$). By increasing the value of $\theta$, these two lattices become detached from each other and moiré patterns emerge. The resulting moiré patterns of two triangular lattices with twist angles $\theta = 0, 2.5, 5$, and $7.5$ are shown in Figure 1.

Critical current densities are calculated for the emerged moiré patterns with different $\theta$ values, ranging from zero to 60 degrees. As a result of the inherent symmetry of the triangular lattice, rotating the lattice by 60 degrees brings it back to its original state. This means that a twist angle of $\theta = 60$ degrees is equivalent to having no twist at all ($\theta = 0$). When $\theta = 0$, the pinning sites in two triangular lattices overlap, resulting in a doubled pinning force. However, upon exiting from the $\theta = 0$ state, the pinning sites no longer overlap and each pinning site has a pinning force that is half of its value in comparison with the $\theta = 0$ state.

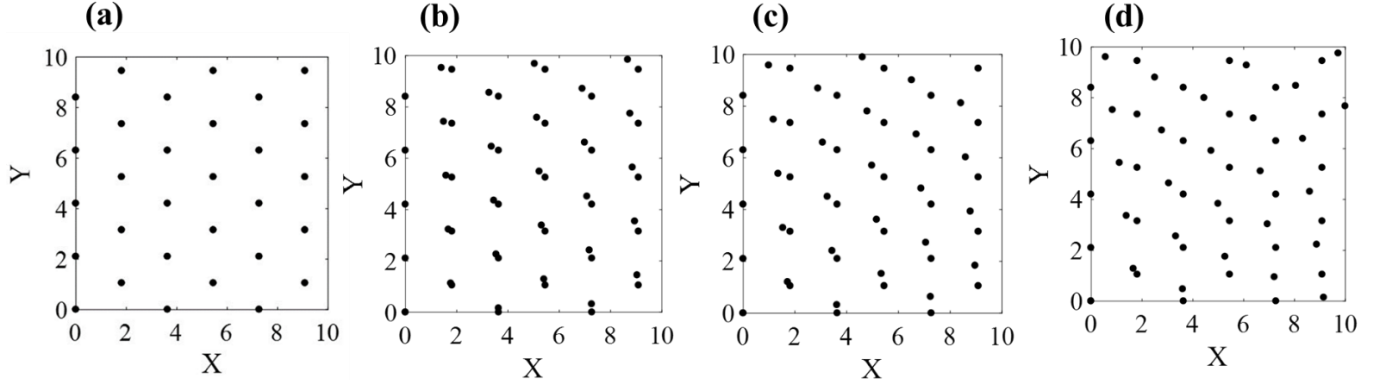

**Figure1**: The moiré pinning array patterns for twist angles (a) $\theta = 0$, (b) $\theta = 2.5$, (c) $\theta = 5$, and (d) $\theta = 7.5$. X and Y are in units of $\lambda$.

## III- Results and Discussion

Figure 2 illustrates the normalized $J_c$ versus the number of vortices for $\theta = 0$ and for pinning forces equal to 0.5, 1, 2, and 3. As expected, for $f_p = 3$, the critical current density is larger than the other cases. Pinning centers act as barriers hindering the movement of vortices. So, stronger pinning forces suppress the vortex motion more effectively, which consequently leads to an increase in the critical current density.

The figure shows that when $\theta = 0$, in addition to the initial peak, there are other peaks for $f_p=2$ and $f_p=3$. The appearance of these peaks has captured the interest of scientists due to their intriguing nature and wide range of applications [68-71].

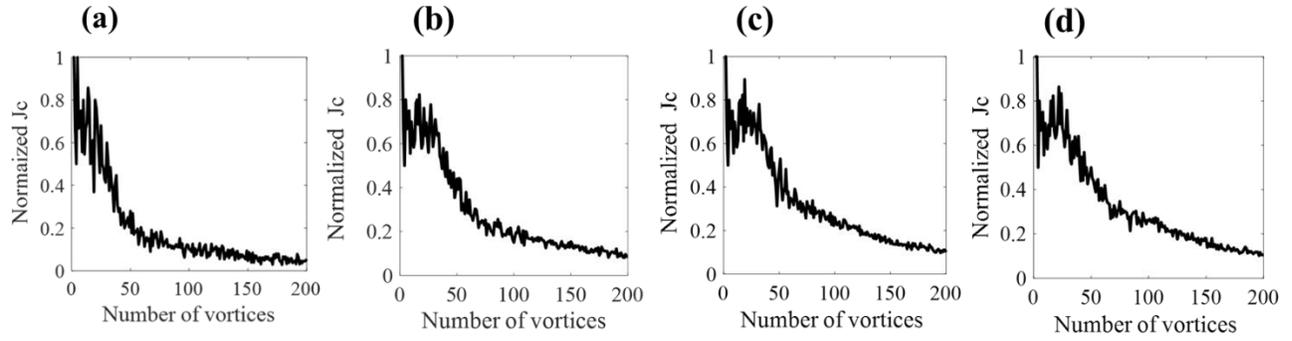

**Figure 2**: Normalized critical current density, $J_c$, versus the number of vortices for (a) $\theta = 0, f_p=0.5$, (b) $\theta = 0, f_p =1$, (c) $\theta = 0, f_p =2$, and (d) $\theta = 0, f_p =3$.

In Figure 3, diagrams of the normalized $J_c$ versus twist angles, $\theta$, are depicted for $f_p = 0.5, 1, 2,$ and 3, respectively, for high, intermediate, and low magnetic fields. High fields refer to the range of vortex numbers between 180 and 200, intermediate fields span the range between 90 and 110, and low fields indicate the range between 40 and 60 number of vortices. The pinning sites are located at the points of two triangular lattices that are co-centered and twisted relative to each other by the twist angle $\theta$.

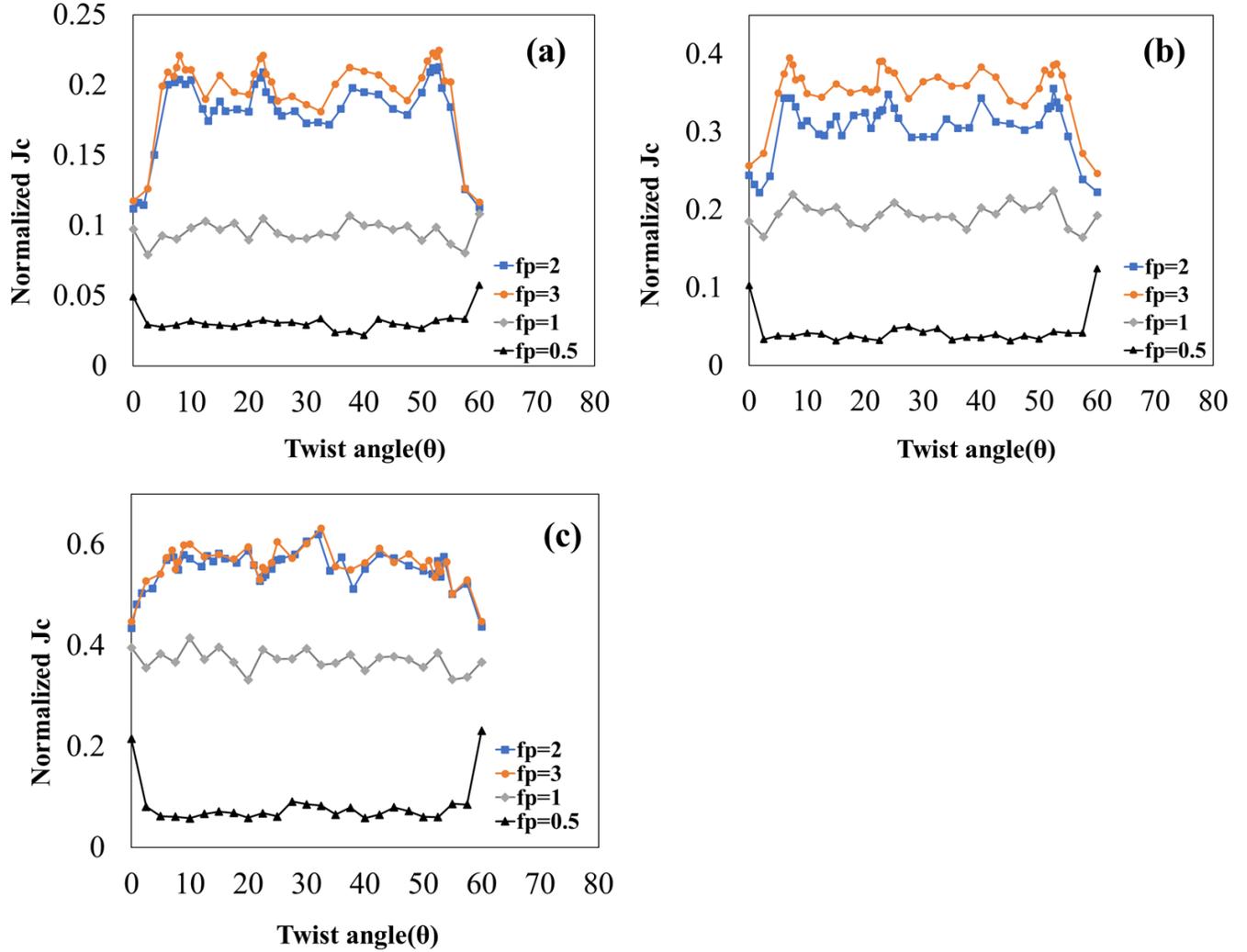

**Figure 3**: Normalized critical current density, $J_c$, as a function of twist angle $\theta$, for high (a), intermediate (b), and low (c) magnetic fields.

Figure 3(a), illustrates the normalized $J_c$ versus $\theta$ for high magnetic fields. From this figure, it can be seen that for $f_p = 3$, the normalized $J_c$ has the largest value for all twist angles in comparison to other pinning force strengths. Also, for $f_p = 3$, a noticeable increase in the critical current density can be observed by increasing the value of $\theta$ from zero. Furthermore, we observe that for the $\theta$ values of 8, 22.5, 37.5, and 53 degrees, the critical current density is maximized, with the absolute maximum value occurring at $\theta = 53$ degrees. The behavior of the critical current density for $f_p = 2$ is similar to that of $f_p = 3$, except for a slight decrease in its value for most of the $\theta$ values. For

$f_p = 1$, the critical current density shows a significant decrease compared to the $f_p = 2$ and $f_p = 3$ cases and does not change with the variations in $\theta$ significantly, and only minor fluctuations are observed.

For $f_p = 0.5$, the critical current densities are smaller compared to the $f_p = 1$ case. Increasing the value of $\theta$ from zero leads to a reduction in the critical current density, similar to the $f_p = 1$ situation. As $\theta$ increases, the normalized $J_c$ remains almost constant and at $\theta = 60$, returns to its initial value.

In Figure 3(b), the critical current density is depicted as a function of $\theta$ for four different values of $f_p$ equal to 0.5, 1, 2, and 3, in intermediate magnetic fields. Within this range of magnetic field strengths, $J_c$ demonstrates a pattern similar to that observed in high fields. Comparing the critical current density plots for intermediate and high magnetic fields, we observe a noticeable enhancement in the critical current density, going from the $f_p = 2$ to the $f_p = 3$ case.

Figure 3(c) illustrates the relationship between the normalized critical current density and the twist angle for four values of $f_p$ equal to 0.5, 1, 2, and 3, in low magnetic fields. Within this range of magnetic field strengths, there is no noticeable difference between $f_p=2$ and $f_p=3$ values in the normalized $J_c$ diagrams.

Determination of the normalized critical current density involves a competition between two key factors; the regularity of the lattice and the degree of the overlap between the pinning sites. When $\theta$ is equal to zero, the pinning centers exhibit an initial arrangement characterized by triangular ordering. Comparing the cases $\theta = 0$ and $\theta \neq 0$, it can be observed that in the $\theta = 0$ case, the pinning force is doubled due to the overlapping of two lattices. However, the number of pinning sites is also halved relative to the non-zero $\theta$ case.

On the other hand, for θ = 0, there is a triangular ordering, whereas for non-zero θ, short-range ordering is lost. Consequently, when θ is non-zero, by the alteration of the initial ordering of pinning centers, a noticeable increase in the critical current density can be observed for pinning forces equal to 2 and 3.

When the vortex-pinning forces are greater than the vortex-vortex forces, the impact of the overlapping between pinning centers becomes more important than the lattice regularity. Therefore, as the angle θ increases, the critical current density increases due to the doubling of the number of pinning sites. The reason for this is the larger magnitude of the vortex-pinning forces, compared to the vortex-vortex forces.

As a general trend, as θ increases, for pinning forces greater than one, where the vortex-pinning forces surpass the vortex-vortex forces, the effect of increased number of pinning sites becomes more significant compared to the increased disorder in the spatial distribution of the pinning centers. Hence, increasing the number of pinning centers leads to an enhancement in the critical current density.

For $f_p = 1$, the vortex-pinning force is equal to the vortex-vortex force. Therefore, with an increase in θ, no significant change in the critical current density is observed. For $f_p = 0.5$, since the vortex-pinning forces are smaller than the vortex-vortex forces, with an increase in the value of θ from zero, the impact of lattice irregularity becomes more prominent compared to the increased number of pinning centers. This results in a decrease in the normalized critical current density.

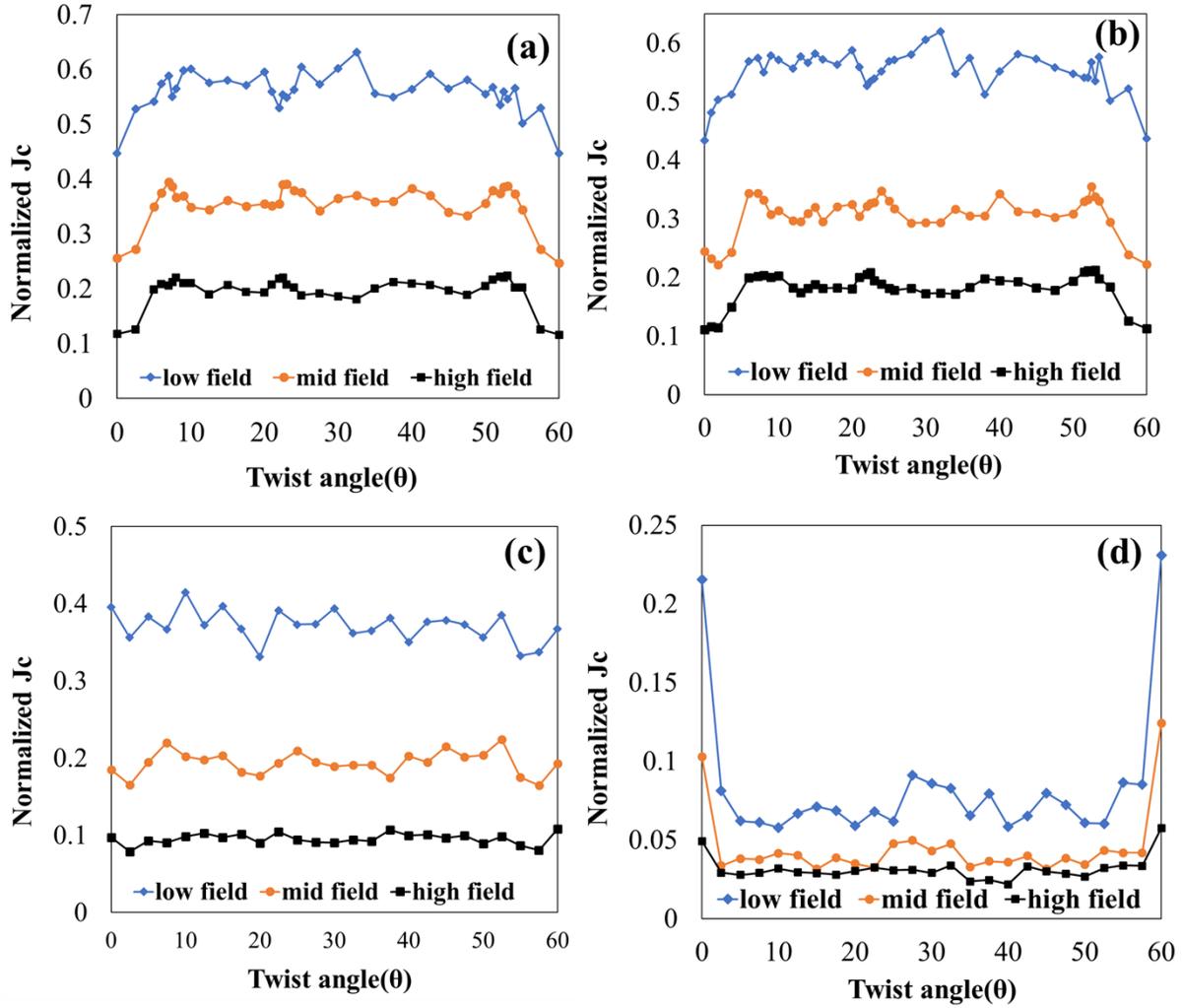

**Figure 4**: Normalized critical current density, $J_c$, as a function of twist angle $\theta$, for $f_p = 3$ (a), $f_p = 2$ (b), $f_p = 1$ (c), and $f_p = 0.5$ (d), in high, intermediate, and low magnetic fields.

In Figures 4(a) to (d), the normalized critical current densities are depicted as a function of the twist angle θ, respectively for pinning forces $f_p$ equal to 0.5, 1, 2, and 3, respectively in high, intermediate, and low magnetic fields. In Figure 4(a), when $f_p = 3$, the maximum critical current density occurs approximately at the same twist angle for the both high and the intermediate fields. An increase in $J_c$ at specific twist angles has also been reported in previous research [72].

In Figure 4(b), the critical current density as a function of angle is shown for $f_p = 2$ in the low, intermediate, and high magnetic fields. As it can be seen, the behavior of $J_c$ resembles that of the $f_p = 3$ case. In Figure 4(c), when $f_p = 1$, $J_c$ remains almost constant for all magnetic field strengths, despite small fluctuations.

Figure 4(d) shows the relationship between the critical current density and twist angle for the pinning force $f_p = 0.5$. In this case, as we exit from the $\theta = 0$ state, the critical current density decreases, initially, then shows fluctuations at higher twist angles, and finally returns to its initial value at $\theta = 60$ degrees. By increasing the pinning force, the difference in the critical current densities corresponding to different magnetic fields decreases.

Comparing the critical current densities at $\theta = 0$ with larger twist angles for $f_p = 0.5$, we observe that a smaller vortex-pinning force leads to a smaller critical current density. For $f_p=2$ and $f_p=3$, the critical current density shows an increasing trend by increasing the value of $\theta$, because, the vortex-pinning forces become larger than the vortex-vortex forces. For $f_p = 1$, the critical current density remains almost constant at all $\theta$ values, with minor fluctuations.

## IV- Conclusion

In summary, we have investigated the normalized critical current density in a superconducting tape with pinning centers forming moiré patterns, obtained from twisting two triangular pinning lattices relative to each other. We have studied the effect of twist angle, different pinning forces, and different magnetic field strengths on $J_c$. Results show that for pinning forces greater than one, because of the larger vortex-pinning forces in comparison with the vortex-vortex forces, the critical current density significantly increases. In addition, the specific twist angles for which the critical current density reaches its maximum values are inspected. When the vortex-pinning forces are smaller than the vortex-vortex forces, and for pinning forces smaller than one, increasing the twist angle results in a decrease in the critical current density. This decrease can be attributed to the lattice ordering and to the pinning sites overlapping.